\documentclass[traditabstract]{aa} 
\usepackage{graphicx}
\usepackage{txfonts}
\begin{document}
   \title{U\,Sco 2010 outburst: a new understanding of the binary \\ accretion disk and the secondary star
\thanks{Based on observations carried out at the European Southern Observatory,
        under programs 084.A-9003 \& 284.D-5041.}}

   \subtitle{}

   \author{E. Mason\inst{1}
          \and
          A. Ederoclite\inst{2}
	  \and
	  R. E. Williams\inst{1}
	  \and
	  M. Della Valle\inst{3}
	  \and	
	  J. Setiawan\inst{4}
          }

   \institute{Space Telescope Science Institute, Baltimore, MD 21218, U.S.A.\\
              \email{emason@stci.edu}
         \and
	  Centro de Estudios de Física del Cosmos de Aragón, Teruel, Spain
         \and
	  INAF, Osservatorio Astronomico di Capodimonte, Napoli, Italy
         \and
	  Embassy of the Republic of Indonesia, Berlin, Germany
             }

   \date{Received May 2012; accepted ...}

 
  \abstract
   {We present optical and NIR spectroscopic observations of U\,Sco 2010 outburst. From the analysis of lines profiles we identify a broad and a narrow component and show that the latter originates from the reforming accretion disk. We show that the accretion resumes shortly after the outburst, on day +8, roughly when the super-soft (SSS) X-ray phase starts. Consequently U\,Sco SSS phase is fueled (in part or fully) by accretion and should not be used to estimate $m_{\mathrm{rem}}$, the mass of accreted material which has not been ejected during the outburst. In addition, most of the He emission lines, and the He\,{\sc ii} lies in particular, form in the accretion flow/disk within the binary and are optically thick, thus preventing an accurate abundance determination. 

   A late spectrum taken in quiescence and during eclipse shows Ca\,{\sc ii}\,H\&K, the G-band and Mg\,{\sc i}b absorption from the secondary star. However, no other significant secondary star features have been observed at longer wavelengths and in the NIR band. }
   {}
   {}
   {}
   {}


   \maketitle
%

\section{Introduction}

U\,Sco is a well know recurrent nova (RN - see Bode and Evans 2008, and/or Warner 1995 for review) which has been observed in outburst in 1863, 1906, 1917, 1936, 1945, 1969, 1979, 1987, 1999 and 2010 (Schaefer et al. 2010). It is also the prototype of the  RNe subclass, which is characterized by 1) very fast outburst decline and evolution, 2) orbital periods of the order of one day, which point to an evolved secondary star, and 3) no development of forbidden emission lines (i.e. a nebular spectrum) from the ejecta.

The interest in RNe is linked to the fact that they are possible candidates for supernova (SN) type Ia progenitors, due to their massive white dwarf. In addition U\,Sco has been a highly debated object for the nature and composition of its donor star, which are still quite uncertain: past works have often reported anomalously high He abundances and concluded that the donor is possibly a He rich and evolved star (Barlow et al. 1981, Williams et al. 1981, Evans et al. 2001, but see also Iijima 2002, Anupama and Dewangan 2000, and Maxwell et al. 2012); while multicolor photometry and spectral analysis have identified it with a G0 (Hanes 1985) or K2 type star (Anupama and Dewangan 2000).

The U\,Sco 2010 outburst had an extensive follow-up both from space and ground, and from X-ray to NIR wavelengths (Schaefer et al. 2010, Munari et al. 2010, Schaefer et al. 2011, Banerjee et al. 2010, Yamanaka et al. 2010, Diaz et al. 2010, Kafka and Williams 2010,  Manousakis et al. 2010, Ness et al. 2012, Orio et al. 2010, Osborne et al. 2010). 
The observations of the 2010 outburst have already provided two important results, revising  our current understanding of this RN: i) U\,Sco develops a nebular spectrum (provided it is followed long enough after the outburst) as shown by Diaz et al. (2010) and Mason (2011); and ii) U\,Sco hosts an ONe white dwarf (Mason 2011), whose ultimate fate, in the case of mass accretion, is to collapse in a neutron star rather than explode as SN-Ia (e.g. Nomoto and Kondo 1991). 

In this paper we present a series of spectra collected across the U\,Sco 2010 outburst with FEROS and X-Shooter and reveal new unexpected results about the outburst evolution and the binary system.  


\section{Observations and data reduction}

\begin{table*}[t]
\scriptsize
\flushleft
 \begin{minipage}{140mm}
  \caption{Log of the observations}
  \begin{tabular}{@{}lclllccccc@{}}
  \hline
   UT Date   &  mid-HJD  & days after  & orbital & {\it V\,$^{\star\star}$} & Instrument &  \multicolumn{3}{c}{ setup (slit$^\dagger$; R; binning$^{\dagger\dagger}$; exptime) } & sky  \\
 (start)  & (-245500.) & outburst &  phase$^\star$ & (mag) &  & {\it UVB} & {\it VIS} & {\it NIR}  & transparency/seeing ($^{\prime\prime}$) \\
        &  &  &  &  &  &  &  & &  \\
 
\hline
29-Jan-2010 & 225.8837 & 1.2 & 0.55 & 9.30 & FEROS  & & 1.8; 48000; 1$\times$1; 900s &  & CLR/PHO; NA \\
30-Jan-2010 & 226.8879 & 2.2 & 0.37 & 9.93 &   FEROS & &  1.8; 48000; 1$\times$1; 720s &  & THN/TCK; NA \\
02-Feb-2010 & 229.8634 & 5  & 0.79 & 11.33 &   FEROS &  & 1.8;  48000; 1$\times$1; 240s$\times$2+60s$\times$6 &  & THN/TCK; NA \\
05-Feb-2010 & 232.8796 & 8  & 0.24 & 12.9 &  FEROS & & 1.8; 48000; 1$\times$1; 280s$\times$6 &  & NA; NA \\
07-Feb-2010 & 234.8751 & 10 & 0.86 & 13.7 &  FEROS &  & 1.8; 48000; 1$\times$1; 300s$\times$3 &  & NA; NA \\
14-Feb-2010 & 241.8630 & 17 & 0.54 & 14.28 & X-Shooter&  0.8; 6200; 1$\times$1; 120s$\times$2 & 0.7; 11000; 100,1$\times$1; 190s$\times$2  & 0.6; 8100; 270s$\times$2 &  CLR; 0.9-3.0\\
24-Feb-2010 & 251.8696 & 27 & 0.67 & 14.60 & X-Shooter & 0.8; 6200; 1$\times$1; 180s$\times$2 & 0.7; 11000; 1$\times$1; 250s$\times$2 & 0.6; 8100; 113s$\times$6 &  THN; 0.6-3.5\\
15-Mar-2010 & 270.8176 & 46 & 0.07 & 17.61 & X-Shooter & 0.8; 6200; 1$\times$1; 300s$\times$4 & 0.7; 11000; 1$\times$1; 370s.4 & 0.6; 8100; 153s$\times$12 & CLR;  0.7\\
11-Apr-2010 & 297.8250 & 73 & 0.02 & 18.28 & X-Shooter & 1.0; 2550; 1$\times$2; 600s$\times$4 & 0.9; 4400; 1$\times$2; 638s$\times$4 & 0.6; 8100; 227s$\times$12 & PHO; 0.6 \\
12-May-2010 & 328.7723 & 104 & 0.17 & 18:   & X-Shooter & 1.0; 2550; 1$\times$2; 1030s$\times$2 & 0.9; 4400; 1$\times$2; 1068s$\times$2  & 0.6; 8100; 300s$\times$6 & CLR; 0.6 \\
05-Jul-2010 & 382.7130 & 125 & 0.00 & 19.16 & X-Shooter & 1.0; 2550; 1$\times$2; 1030s$\times$2 & 0.9; 4400; 1$\times$2; 1068s$\times$2 & 0.6; 8100; 300s$\times$6 & CLR; 0.7-1.7\\
\hline
\end{tabular}
$^\star$ The spectra have been phased using the ephemeris in Schaefer et al. (2011). \\
$^{\star\star}$ Courtesy of Brad Schaefer.\\
$^\dagger$ In the case of FEROS, 1.8 is the fiber diameter on sky, in  arc-sec.

$^{\dagger\dagger}$  In the case of X-Shooter the binning is in the dispersion direction. 
\end{minipage}
\end{table*}

The collection of data presented in this paper were obtained with the ESO 2.2m+FEROS spectrograph for the maximum and early decline epochs and with the ESO VLT+X-Shooter for the later epochs. Table\,1 reports the journal of observations. Both FEROS and X-Shooter are cross dispersed echelle spectrographs with fixed spectral format. FEROS is a fiber-fed instrument which delivers optical ($\lambda$-range $\sim$3800-9000\,\AA) spectra of resolution R=48000; while X-Shooter is a medium resolution spectrograph (R$\leq$18000), which covers the whole optical+NIR wavelength range in a single shot.   
The data were reduced using each instrument pipeline, though the X-Shooter spectra were extracted and merged using IRAF routines (see Mason 2011 for details). 

Flux calibration was performed using IRAF and user scripts.  We used the spectrophotometric standard stars HR\,4963 (FEROS) and EG\,274 (X-Shooter) observed on Feb\,7 and May\,12 2010, respectively. Before flux calibration the VIS and NIR arm X-Shooter spectra where corrected for telluric absorptions using the telluric star Hip\,82254. 
However, we used normalized spectra (i.e. not corrected for telluric absorptions, nor flux calibrated) when analyzing the line profile to avoid introducing possible spurious features. A montage of normalized spectra is presented in Fig.\,A.1 through A.3. 

\section{The broad emission lines from the ejecta and their evolution}

Both optical and NIR spectral evolution of 2010 U\,Sco outburst have already been presented by Diaz et al. (2010), Yamanaka et al. (2010), Banerjee et al. (2010) and Maxwell et al. (2012) and we do not intend to repeat their analysis here. However, as 
our sequence of spectra covers all the major phases of the outburst it is of interest to match the spectral evolution of the nova to the light curve phases that have been identified by Schaefer et al. (2011). 
Fig.\,1 shows U\,Sco light curve, the light-curve phases identified by Schaefer and the epochs of our spectroscopic observations.

\begin{figure}
\centering
\includegraphics[width=8.5cm, angle=0]{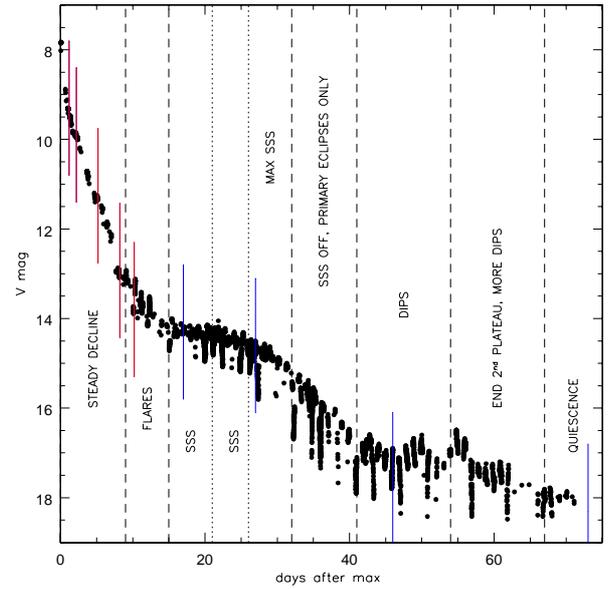}
\caption{U\,Sco V-band light curve (courtesy of Brad Schaefer). The vertical black dashed lines marks the evolutionary phases identified by Schaefer et al. (2011), while the red and blue vertical segments marks the epoch of our FEROS and X-Shooter spectroscopic observations, respectively. Each of the light curve evolutionary phases has been marked in the figure itself. More details can be found in Schaefer et al. (2011).}
\label{LC}%
\end{figure}

  \begin{figure}[h!]
   \centering
   \includegraphics[width=8cm]{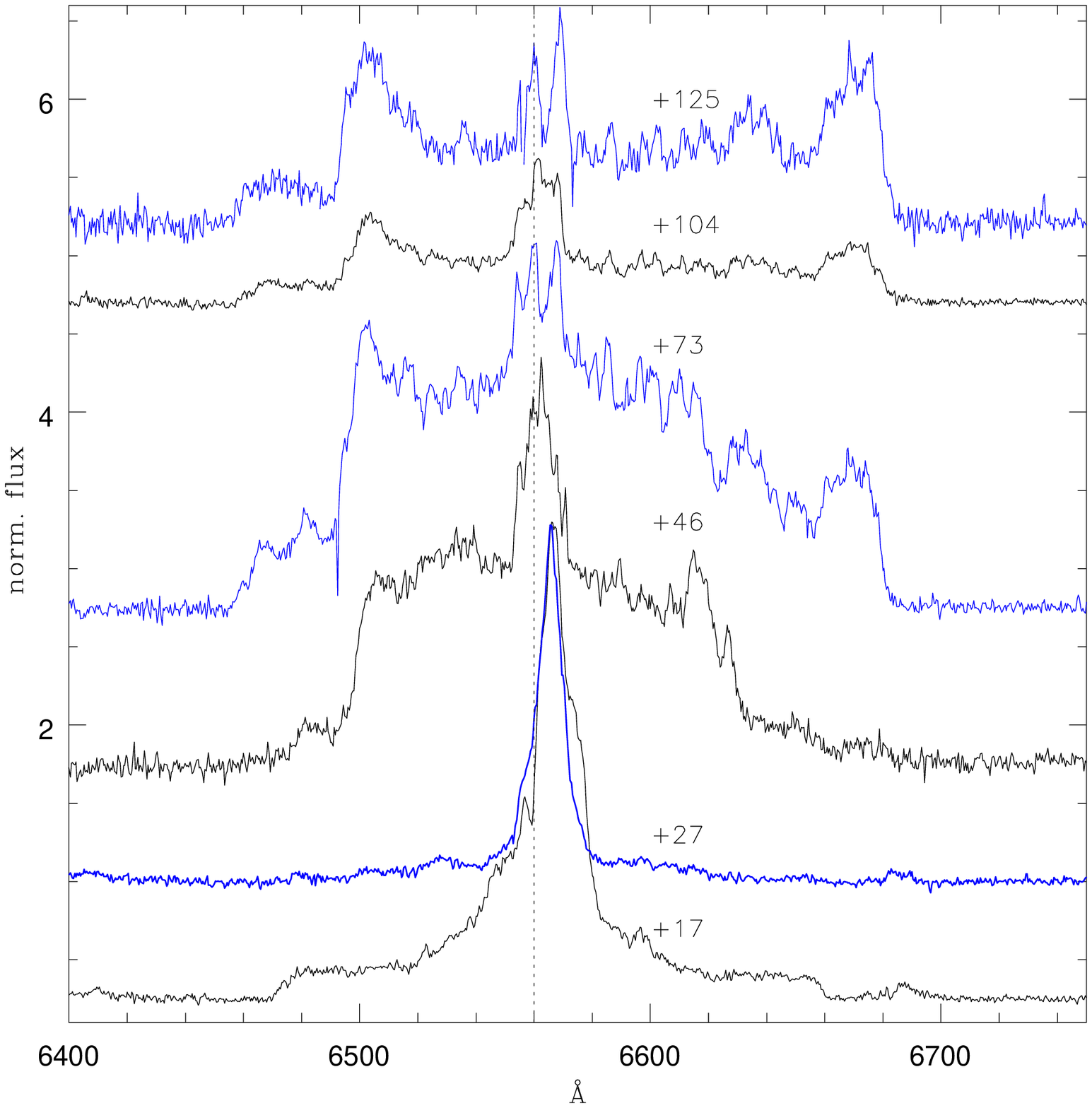}
   \includegraphics[width=8cm]{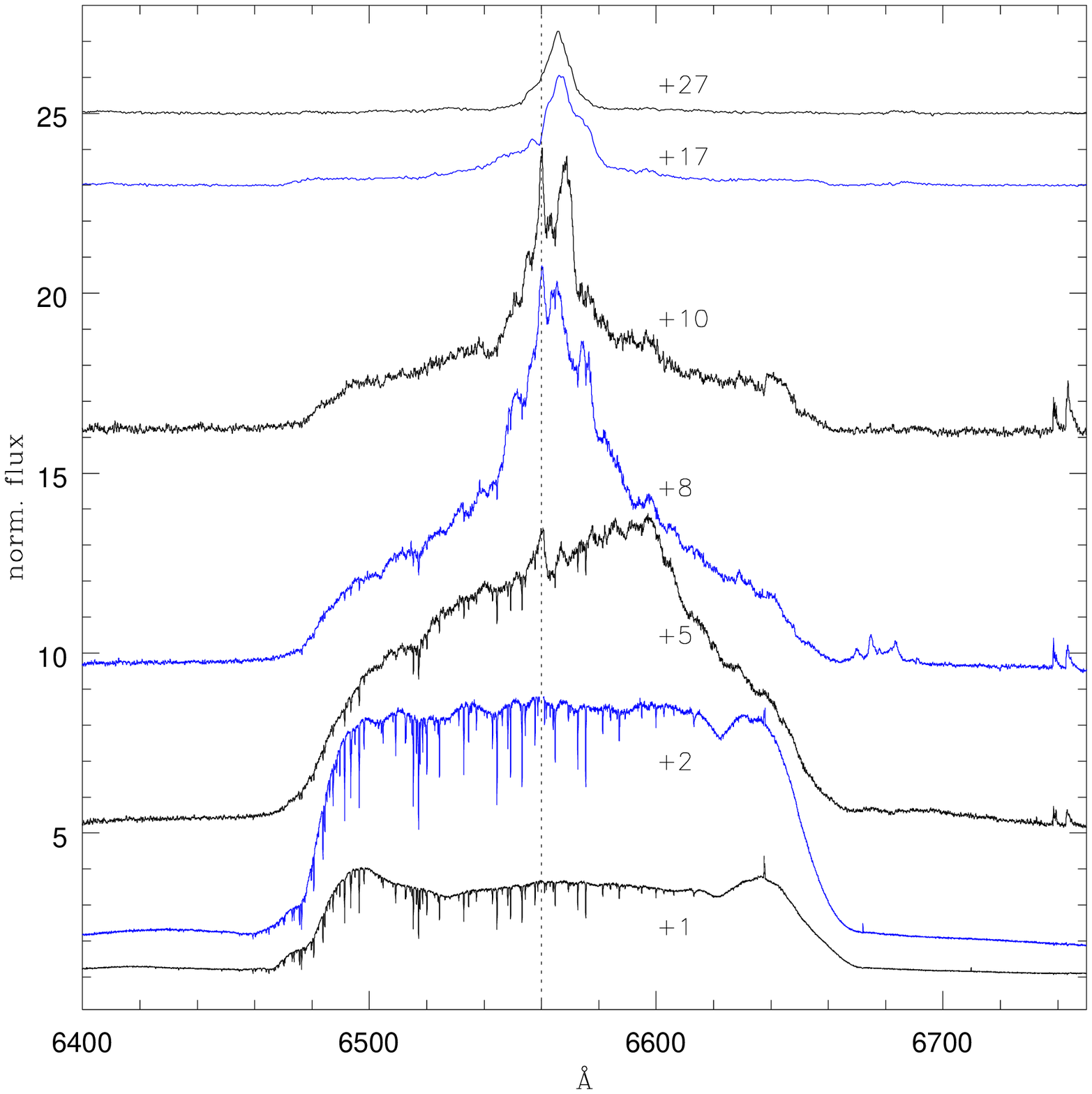}
   \caption{Bottom Panel: montage of the H$\alpha$ profile evolution during the first 7 epochs of observations. Time goes from bottom to top, and the first 5 spectra were taken with the 2.2m+FEROS. Top panel: montage of the H$\alpha$ profile evolution during the last 6 epochs which were taken with the VLT+X-Shooter. Note that epochs 6 and 7 appear in both panels. The sequence has been presented in two separate panels for clarity: the intensity of the broad component decreased by a factor $\sim$65 from epoch 2 (at maximum light) to epoch 7 (see Fig.\,3). The dotted vertical line marks the H$\alpha$ rest wavelength. The numbers near each spectrum mark the epoch of the spectrum itself in days after maximum.  }
              \label{}%
    \end{figure}

At maximum light (days +1 \& +2), U\,Sco appears as a typical He/N fast nova with extremely broad  rectangular emission lines (FWHM$\sim$8000\,km/s; Fig.\,A.1-A.3 and Fig.2). 
Rectangular or squared profiles are associated with optically thin lines emitted by an expanding shell (Williams 1992). However, U\,Sco ejecta most likely consists of two shells expanding with different velocity. This is clear in the late epochs spectra ($\geq$ +46 days) where the H$\alpha$ and the nebular emission lines (e.g. [OIII]$\lambda$4363) display a 6000\,km/s wide component superposed on the 8000\,km/s component. We interpret the spectrum taken on day +5 in the context of two shells, which shows emission lines of a roughly triangular profile.  Their triangular top is the emerging and partially self-absorbed ``slower'' shell. 
The narrow peak that appears from day +8 on, has a different origin which we discuss in Sections\,4 and 6.

\begin{figure}[h!]
\centering
\includegraphics[width=8.5cm, angle=0]{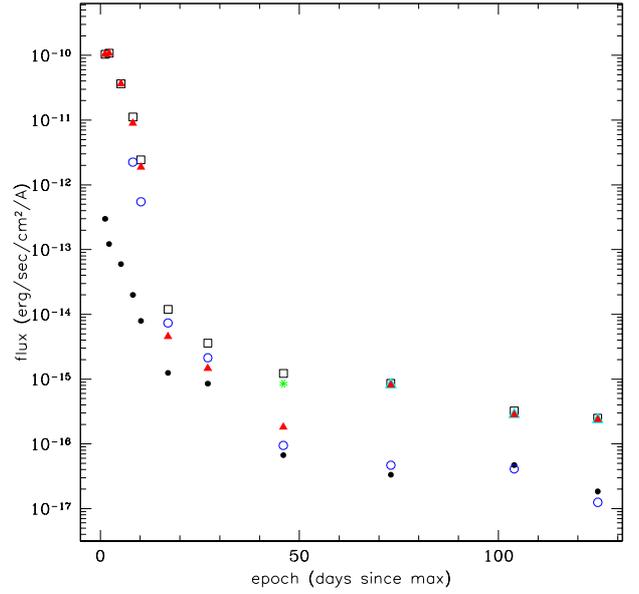}
\caption{The decline of H$\alpha$ emission and its continuum. The black filled circles represent the continuum, black empty squares represent the global H$\alpha$ flux; while the red triangles represent the flux of the 8000\,km/s broad component, and where they are bordered in cyan (last 3 epochs), the major contribution to the line flux is from the [N\,{\sc ii}] doublet. The green asterisk marks the flux of the 6000\,km/s broad component, while the blue empty circles represent the narrow ``disk'' component.  }
\label{LC}%
\end{figure}

During the early decline,  
U\,Sco V-band continuum fades rapidly until day +12, when the X-ray super-soft source (SSS) phase begins (Schlegel et al. 2010). At this time, the optical continuum outshines the ejecta emission lines, which have faded more rapidly than the continuum (see Fig.3). The broad emission lines seem to disappear on day +27 when the super-soft flux is at maximum (see Schaefer et al. 2011, and Fig.1, A.1 and A.2).   
Once the super-soft source (SSS) has turned off ($\sim$ day +32) and U\,Sco has entered the second plateau ($>$+41 days), the broad emission lines from the ejecta are visible again, with the ''narrower'' component (FWHM$\sim$6000\,km/s) being much stronger than the broad (FWHM$\sim$8000\,km/s; see Fig.\,A.1-A.3). At this time the forbidden nebular transitions  (mainly [O{\sc iii}] and [Ne{\sc v}]) appear and their analysis shows that the two ejecta components have different densities (see Fig.\,4), with the narrow 6000\,km/s component being roughly twice as dense than the broad (8000\,km/s) component. 
The intensity of He\,{\sc i}$\lambda$10830 also indicates a high gas density in the ``narrow'' ejecta component. 
\begin{figure}[h!]
\centering
\includegraphics[width=7cm, angle=0]{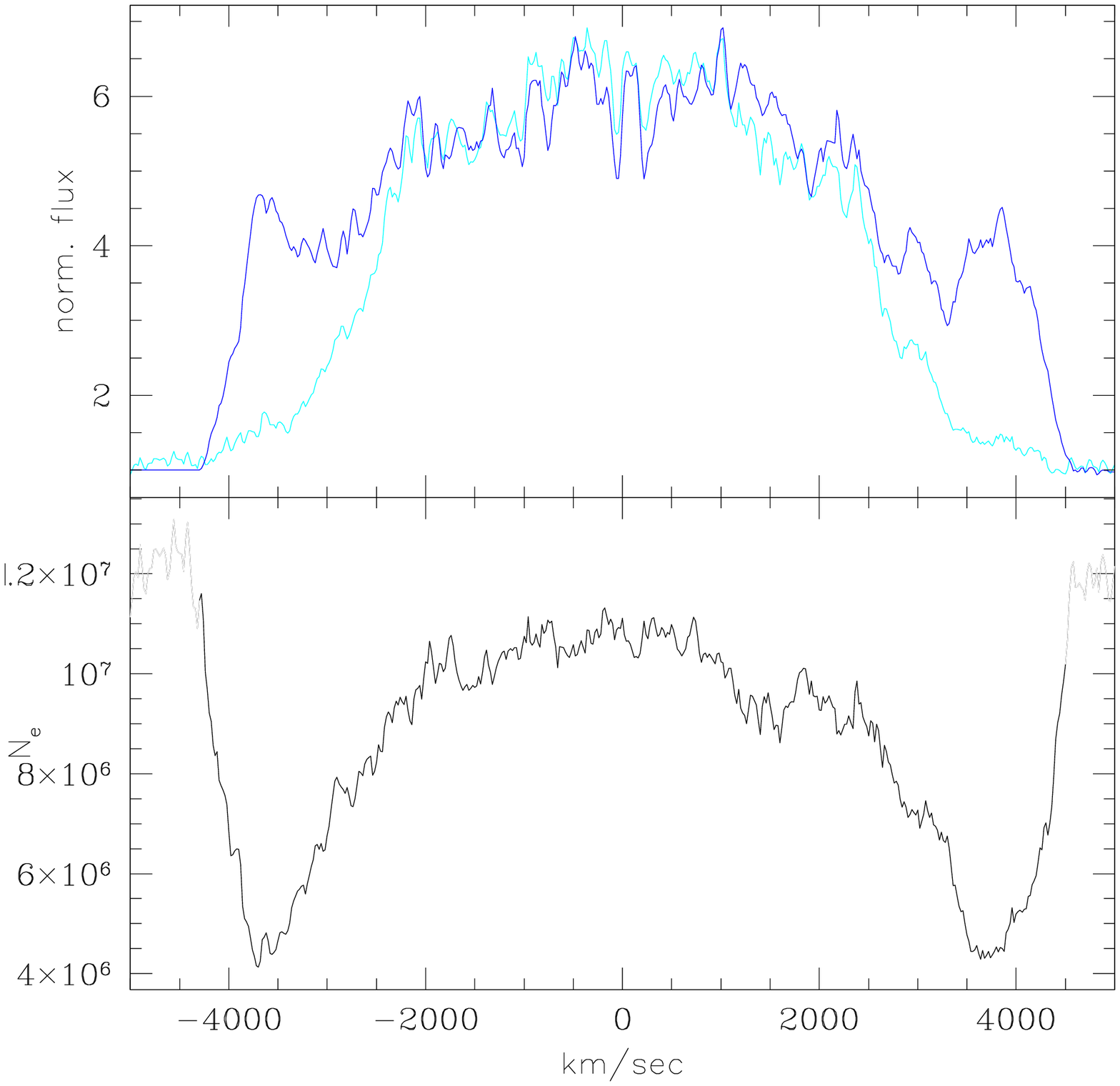}
\caption{The electron density, N$_e$, in U\,Sco ejecta as derived from the [O\,{\sc iii}] flux ratio in the April spectrum. In the top panel the blue and cyan lines represent the profile of the $\lambda$5007 and $\lambda$4363 emission lines, respectively. In the bottom panel the black line shows how the density varies across the emission/ejecta: assuming T$_e$=12000\,K (see Mason 2011 for details), the high velocity wings ($\sim\pm$4000\,km/s) have a density of $\sim$5$\times$10$^6$\,e$^-$/cm$^3$; while the ``low velocity'' component ($\sim\pm$3000\,km/s) has density twice as high, i.e. $\sim$1$\times$10$^7$\,e$^-$/cm$^3$. Note that the density values beyond $\sim$4000-4500\,km/s are spurious due to the uncertain position of the continuum. The contribution of [O\,{\sc iii}]$\lambda$4959 has been subtracted from the $\lambda$5007 emission, assuming symmetric profiles.  }
\label{shore-like}%
\end{figure}
The lack of other He\,{\sc i} and {\sc ii} lines, in combination with the fact that the SSS has turned off, is not compatible with ionization of the He atoms and recombination emission from cascade down. Instead this can be explained by a gas density that is larger than the critical density: for  N$_e > $N$_{critical}$, the 2$^3$P atomic level gets populated by collisions, producing an enhanced He\,{\sc i}$\lambda$10830 emission (Osterbrock 1989). 
As the shell further expands and cools, the H\,Balmer and the He\,{\sc i} emission lines disappear and become replaced by a fully nebular spectrum (+104 days, on; see Fig.\,A.1-A.3). 

We note that in the spectra at maximum (Jan 29 \& 30) the emission lines have relatively low ionization level and in particular, the contribution of the He\,{\sc i} is minimal: Na\,{\sc i} and Mg\,{\sc ii} dominate the emission features at $\sim$5900 and 4480\,\AA, respectively. The He\,{\sc i} emissions lines become more prominent on day +5 when the shell has significantly expanded and the density decreased. The large relative flux of the O\,{\sc i}\,$\lambda$7773 and $\lambda$8446 emission lines in the same spectra is suggestive of high gas densities (possibly as high as 10$^{12}$\,e$^-$/cm$^3$, Williams in preparation) and of collisions dominating the line formation mechanism. In addition, the hydrogen Balmer series displays at all times a decrement that is larger and flatter than the optically thin recombination cases A and B (see Tables\,A.1 and A.2, and compare with Osterbrock 1989). Note that both in Fig.\,3 and in the Appendix's Tables, the flux of the broad component has been obtained by subtracting the narrow component flux (when present) from the integrated flux of the composite emission. The integrated flux of the narrow component has been measured taking its continuum at the top of the broad emission component. This approach provides a lower and an upper limit to the narrow and broad component flux, respectively. 

\section{The narrow emission component from the disk}

Starting from day +8, the U\,Sco emission line spectrum displays a narrow component having FWHM in the range $\sim$650-1100\,km/s. The narrow component is observed in the H-Balmer series, He\,{\sc i} and {\sc ii}, as well in the N\,{\sc iii} at $\sim$4640 and 4512\,\AA, C\,{\sc iv} at $\sim$5800\,\AA \ and possibly O\,{\sc iii} 5592\AA. The N, C and O emission lines, however, are very short lived and have already disappeared by day +10. The H and He narrow emissions persist across the decline phase and in quiescence. They decrease in flux and width with time. 

Yamanaka et al. (2010), observing the development of the narrow component and monitoring the outburst until day +23, ascribed the narrow emission component to a wind which was growing more symmetric and was decelerating as the emitting region moved closer to the white dwarf. However, thanks to the longer monitoring and the higher spectral resolution we suggest that the narrow component originates from the binary system and not the ejecta. 
First, was it part of the ejecta it should have developed nebular emission lines, as has been the case for the broad components. Second, and most important, the narrow component changes in position and profile with time and orbital phase (Fig.\,5). The radial velocity curves of He\,{\sc ii}$\lambda$4686 and a few other isolated He\,{\sc ii} lines (Fig.\,6) mirror neither the secondary star (thus excluding any strong contribution from irradiation), nor the primary orbital motion, as the maximum radial velocity shift is reached at orbital phase $\sim$0.55. In addition, the radial velocity curve in Fig.6 is not sinusoidal in shape, implying that the line forming region is not in a circular orbit. 
We further note that only He\,{\sc ii}$\lambda$4686 and H$\alpha$ are visible at all orbital phases, while, the weaker lines such as the higher emissions of the H Balmer and Paschen series, are visible only when the secondary star is away from the observer. In our set of spectra, the Balmer and Paschen series can be detected down to $\sim$H17 and P17, respectively, at orbital phases 0.54, 0.67 and 0.17. At these same orbital phases the NIR range of U\,Sco spectrum shows a number of narrow emission from He\,{\sc ii} ($n=6$ and 7). This make it clear that most of the H lines we see in the visible are blended with He emissions: the H-Balmer with the He-Pickering, the H-Paschen with the He-Fowler series. Their radial velocities show a large range of values because of the line blending. However, their average value is consistent with the radial velocity curve in Fig.6. The decrement of the He\,{\sc ii} emissions\footnote{Only the isolated/non-blending lines have been ratioed after having measured their integrated flux.} are larger and flatter than the optically thin case (e.g. Osterbrock 1989), implying optically thick gas (see Table\,A.3). 

\begin{figure}[h!]
\centering
\includegraphics[width=7cm, angle=270]{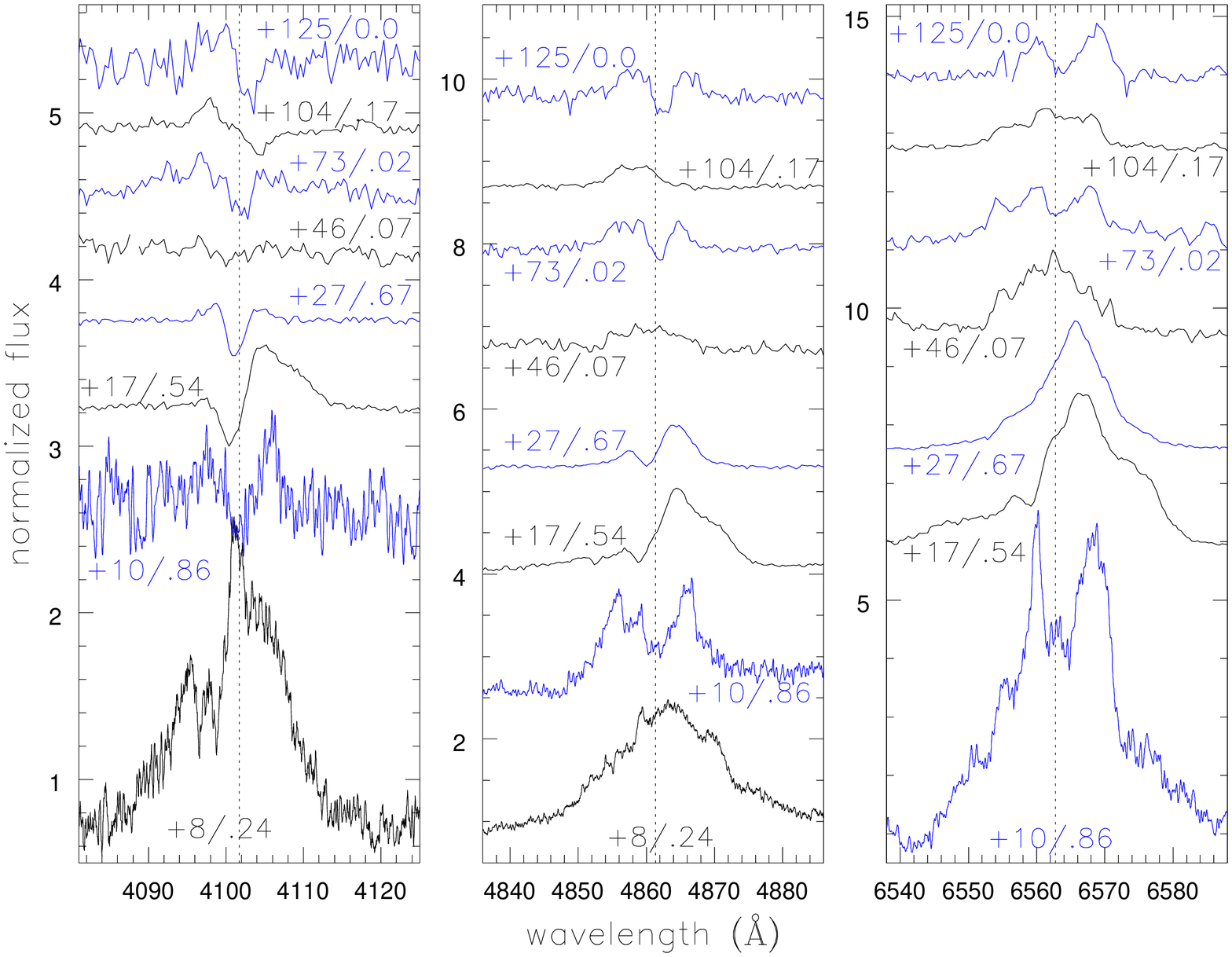}
\includegraphics[width=7cm, angle=270]{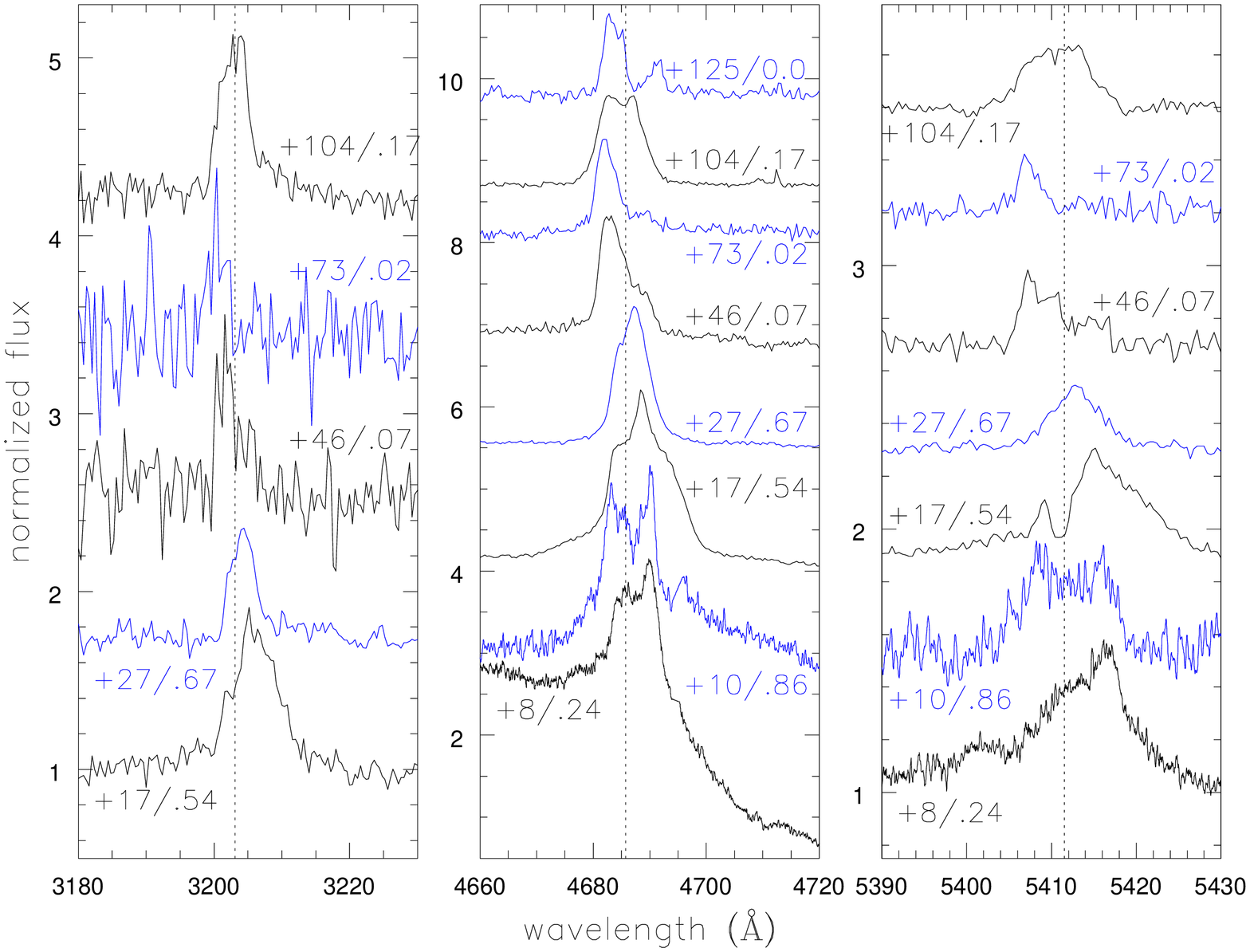}
\caption{Example of U\,Sco the narrow emission component complex profile and its variation with time and orbital phase. The top panel show the Balmer lines H$\delta$, H$\beta$ and H$\alpha$; the bottom panel show the He\,{\sc ii} emission lines at 3203, 4686 and 5411\,\AA. The dashed vertical line marks the rest wavelength positions. }
\label{hEhaNarrow}%
\end{figure}

\begin{figure}[h]
\centering
\includegraphics[width=8.5cm, angle=0]{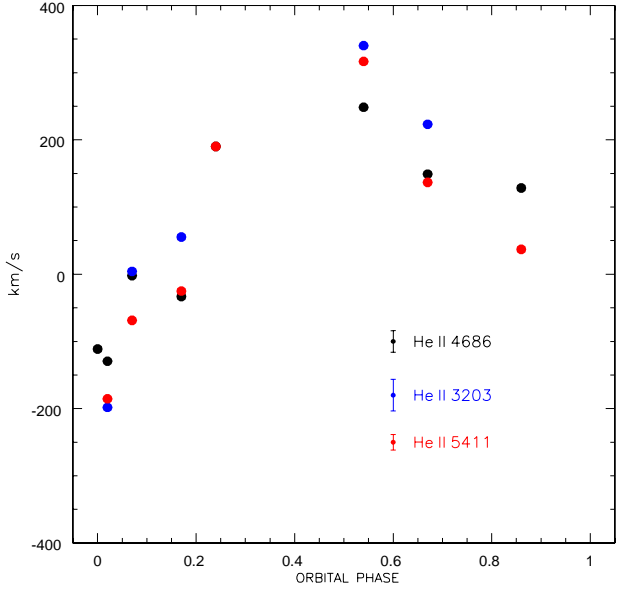}
\caption{Radial velocity curve of the unblended He\,{\sc ii} emission lines derived by measuring the emission line flux barycenter. Different colors represent different lines and the color code is in the figure, together with the uncertainty associated to each line. The error bars plot in the figure correspond to the resolution element (FWHM). Velocities have been corrected for Heliocentric velocity and the zero-wavelength offset discussed in Section\,5.}
\label{rv9}%
\end{figure}

\begin{figure}[h]
\centering
\includegraphics[width=8.5cm, angle=0]{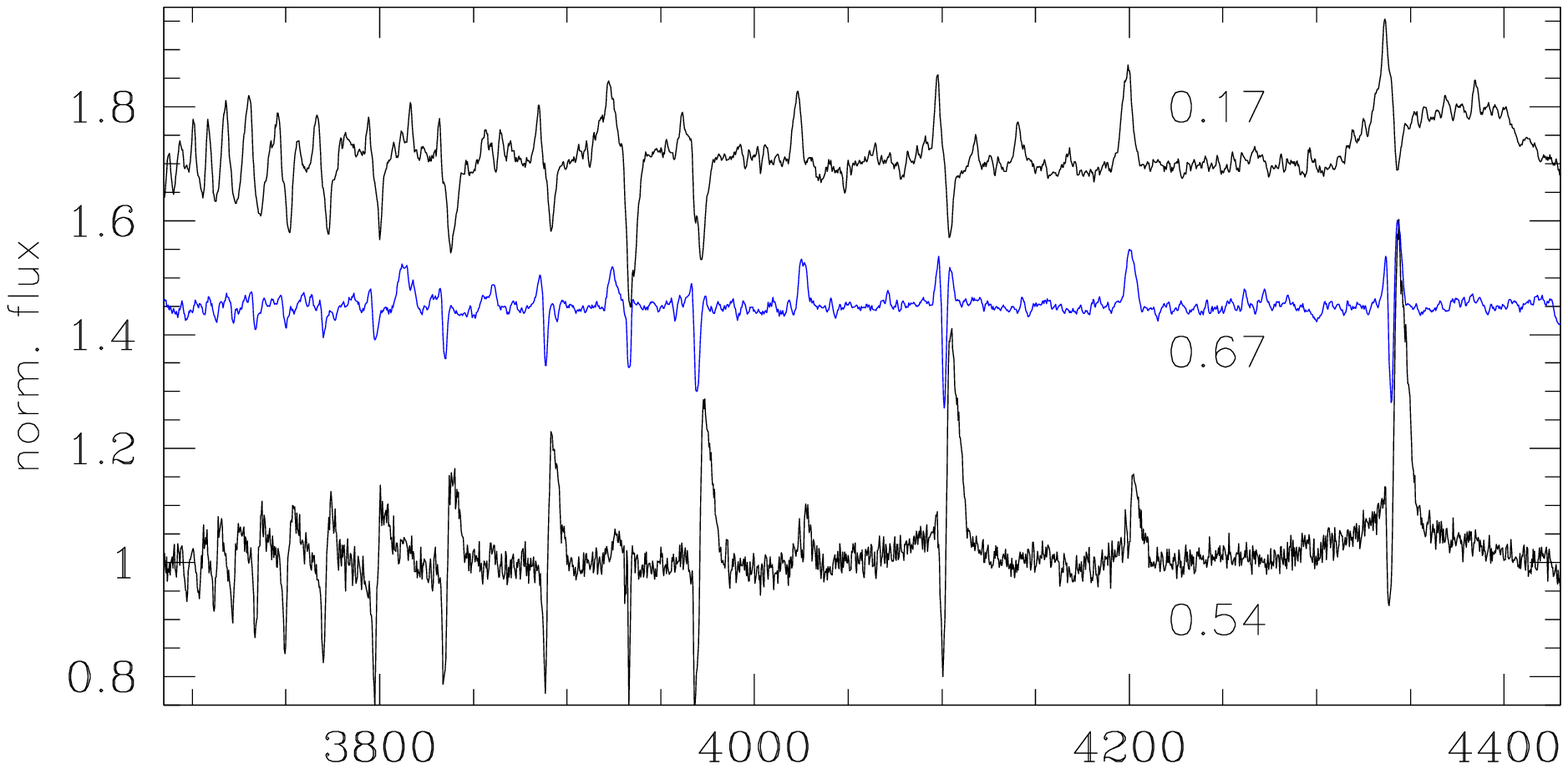}
\includegraphics[width=8.5cm, angle=0]{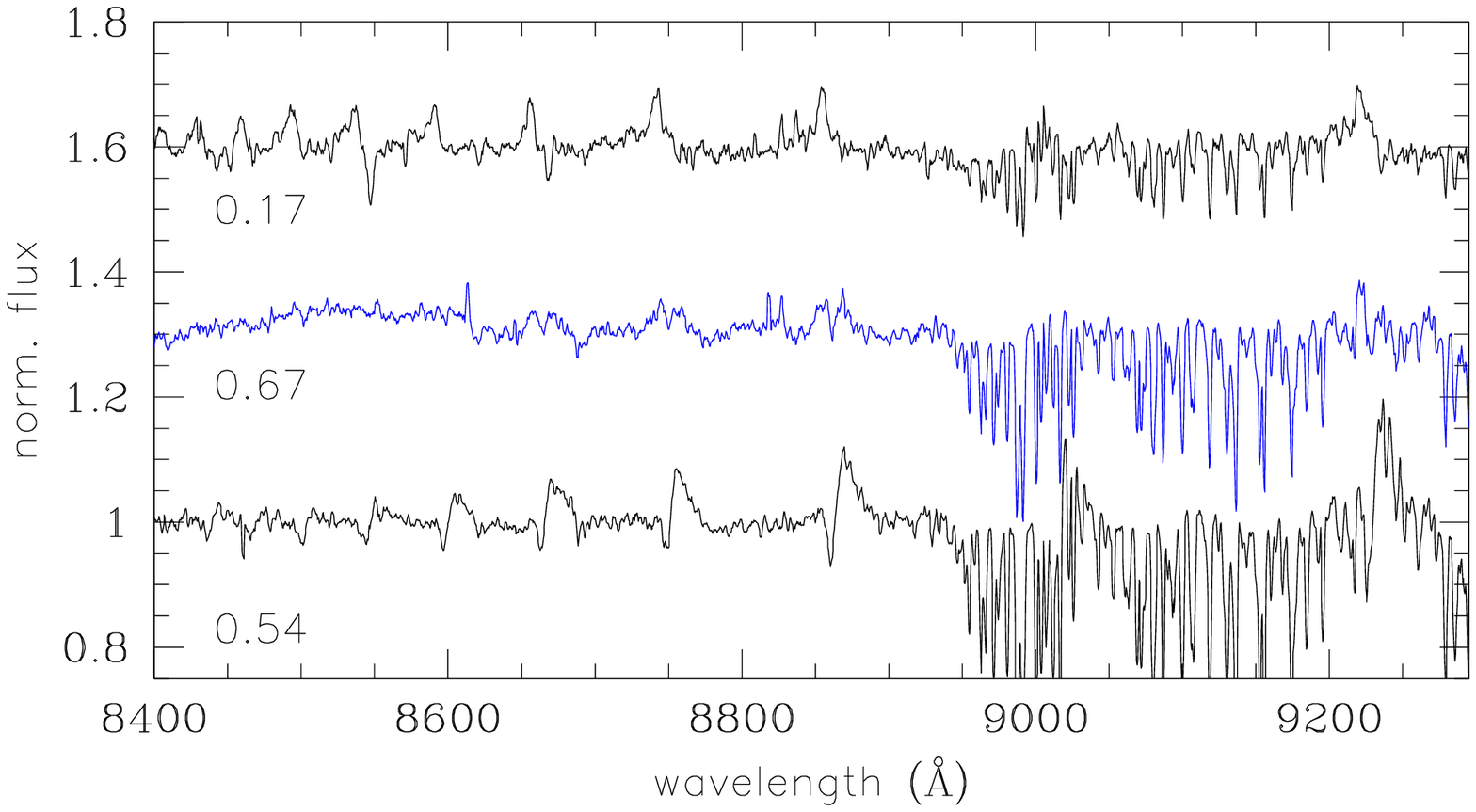}
\caption{The H Balmer (top) and Paschen (bottom) series and their profile variation, as visible in epochs +17, +27 and +104 days. The corresponding orbital phase is marked near each spectrum. Note that all but the bottom spectrum in the top panel have been smoothed with a box function of 5 points. }
\label{Hprofiles10}%
\end{figure}
The narrow hydrogen emission lines also show an absorption component which biases the radial velocity measures and whose relative positions with respect to the emission varies with orbital phase (Fig.\,7). They appear on the blue side of the narrow emission component (forming a P-Cyg like profile) in the +17 days spectrum taken at orbital phase 0.54; on top of the narrow emission line on the +27 days spectrum taken at orbital phase 0.67; and  red-shifted with respect to the narrow emission lines in the +104 days spectrum which was taken at orbital phase 0.17 (Fig.\,7). The radial velocity of the absorption Balmer components are consistent with little or no shift during phase 0.54 and 0.67 ($\sim$-8 to -55 km/s) and have an average velocity shift of +150$\pm$55\,km/s in the 0.17 phase spectrum. Weak absorptions can be observed also in correspondence with some of the He\,{\sc i} and He\,{\sc ii} emission, but they are much less pronounced and more often affecting just the edge of the emission lines. 

The emerging picture is that of the narrow emission and absorption lines forming in different regions, neither of which matches the orbital motion of either star. They must originate within the binary system and most likely in gas which is moving around the primary star but not yet in circular motion around it. 
As much as the radial velocity curve has been monitored over a 3 months period, which is longer than the time needed for an accretion ring to form, circularize and spread into a disk\footnote{This time is about 1 month, assuming a viscosity of $\sim$10$^{15}$\,cm$^2$/s (from Spruit www.mpa-garching.mpg.de/$\sim$henk/pub/disksn.pdf), and adopting a ring of $\sim$5$\times$10$^8$\,cm (see Warner 1995).}, the emission line profiles result from the combination of varying intrinsic velocities and orbital motion. It is therefore difficult to identify a unique line forming region which accounts for both the emission and the absorption components at every phase. 
The spectra with phase 0.54-0.67 were taken during the first plateau and the SSS phase of U\,Sco light curve, when the geometry of the accretion stream possibly matched the initial stream trajectory (e.g. Ness et al. 2011 and their figure 16 in particular). The spectra of phases 0.07-0.17 were taken during the second plateau when the SSS phase has turned off and the accretion disk has reached the quiescent size and geometry (see Schaefer et al. 2011 and their eclipse mapping section). 

By identifying the narrow emission component with the resumed accretion (independently, on its exact location and origin: the accretion flow, the impact region or hot-spot, slingshots between the secondary and the primary stars), we conclude  that mass transfer in U\,Sco restarts as early as Feb\,5, i.e., 8 days after maximum. This is not inconsistent with the start of the SSS phase on Feb\,9 (+12 days after maximum, Schlegel et al. 2012), inasmuch as the Swift observations paused between day +8 and +12 because of lunar constraints and the rise of the soft X-ray emission very likely occurred during the lunar gap (Schlegel et al. 2012). 
Our interpretation is consistent with Worters et al. (2010), who report the resumption of U Sco optical flickering on Feb\,5 UT, i.e., 8 days after outburst. In addition we note that if the SSS phase is fueled by the accretion material it should not start as soon as the mass transfer from the secondary restarts, but some time later, once the accretion ring has spread into a disk reaching and accreting onto the primary. We note also that the expanded white dwarf might significantly reduce those time possibly intercepting the accretion stream directly, similarly to a direct impact accretor. 

Within this picture, if we reanalyze the Schaefer et al. (2011) light curve (see also Fig.8) we can identify the $\sim$0.4 mag drop on day +10.2 as the first partial eclipse of the white dwarf and explain the flares observed between days +10 and +13 as the results of interaction between the secondary and the primary due to the resumed mass transfer. The flares possibly correspond to a transition phase of the nova. The transition phase consists in $\leq$1-2 mag oscillations which might appear when the nova is $\sim$4 mag below the maximum and the outburst light curve changes slope (see e.g. GK\,Per, DK\,Lac, V603\,Aql etc in Bianchini et al. 1992 and Leibowitz et al. 1993). The causes of the transition phase are not yet known (see Phillips and Selby 1977, Bianchini et al. 1992, Shaviv 2001, Retter 2002, for possible scenarios), however, we favor the explanation proposed by Leibowitz et al. (1993) who associate the transition phase with the restart of the accretion process. Observing  early accretion disk eclipses in nova Her\,1992, in addition to the light curve change of slope and ``flares'' as in those novae showing an obvious transition phase, Leibowitz et al. (1993) explained the observations with the accretion disk luminosity overtaking that of the fading nova and the oscillations as a kind of dwarf nova outburst. U\,Sco should not have a fully formed accretion disk at the time of the flares so certainly those cannot be explained with dwarf nova like outbursts. However, it is reasonable to imagine that mass transfer is restored somehow violently because of the changing conditions on both the white dwarf and the secondary star. 

\begin{figure}[h]
\centering
\includegraphics[width=8.5cm, angle=0]{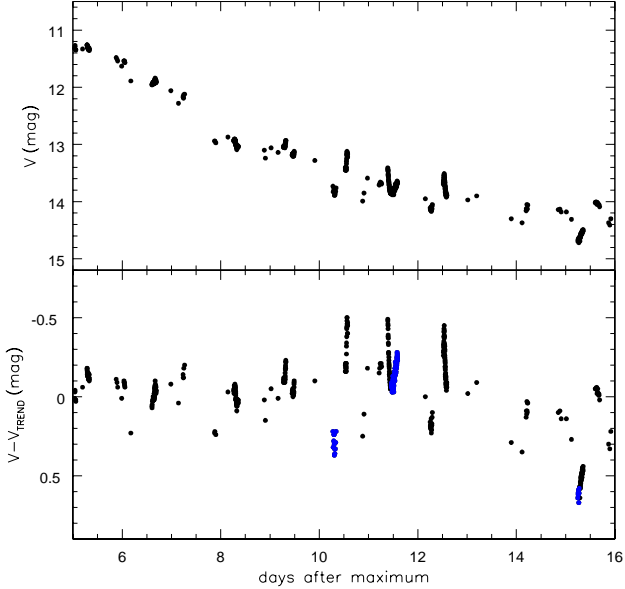}
\caption{Top panel: U\,Sco 2010 outburst light curve from +5 to +16. Bottom panel: U\,Sco light curve in the same period, after subtraction of the decline trend (courtesy of Brad Schaefer). The blue points in the bottom panel marks the observations taken between orbital phase 0.9 and 1, i.e. roughly when the white dwarf eclipse is expected (see Schaefer et al. 2011 for more details).}
\label{brad211}%
\end{figure}

\section{Secondary star features}

The +125\,days spectrum was taken during eclipse in order to observe the secondary star. 
The montage of spectra in Fig.\,A.1-A.3 shows that there are virtually no spectral features but the weak remnant of the [O\,{\sc iii}]\,$\lambda\lambda$4959,5007 and [Ne\,{\sc v}]\,$\lambda\lambda$3345,3425 emissions. However, a closer look at the UVB and VIS arms spectra reveals some secondary star signatures. The Ca\,{\sc ii} H\,\&\,K and the Na\,{\sc i}\,D doublets are evident, but, while the latter originate in the interstellar gas, the former has stellar origin. The Na\,{\sc i} line widths are consistent with non-resolved spectral features, and the $\sim$0.7\,\AA\, shift, which we observe between the first and last epoch of our X-Shooter observations, affects also the telluric spectra (same epochs and very similar pointing directions on sky) and is possibly due to instrumental flexure\footnote{The active flexure compensation system of X-Shooter guarantees the alignment of each spectrograph within the backbone structure, while spectrograph flexure up to 1.2 pixels might be observed moving from Zenith to $z\sim$60$^o$ ($z$ being the Zenith distance). Our observations started at $z\sim$40$^o$ before meridian in February and ended at about the same Zenith distance, after meridian, in July, passing as close as $\sim$7$^o$ to the Zenith. The 0.7\AA\ total shift we observe corresponds to $\sim$1.75 pixels and is roughly consistent (though a bit larger) with the expected 1.6 pixels shift (assuming that the flexure varies linearly with $z$).}. 
The UVB Ca\,{\sc ii} absorptions, on the contrary, are broad (FWHM$\leq$500\,km/s) and can be observed already in the +73 and +104\,days spectra, taken at orbital phase 0.02 and 0.17, respectively, though they are somehow narrower (FWHM$\sim$3-400\,km/s) and weaker.
In addition to the Ca\,{\sc ii} doublet, the Mg\,{\sc i}(2) triplet and the G-band at $\sim$4300\AA\,  should also be ascribed to the secondary star. The Mg\,{\sc i} triplet from the secondary star was first detected by Hanes (1985) and later 
used by Thoroughgood et al. (2001) to measure the radial velocity of the secondary star. 
However, this is possibly the first time that the G-band is detected. Johnston and Kulkarni (1992) could not detect it in their spectra, taken  ~4 yr after the 1987 outburst. 

In our spectra, the Mg\,{\sc i} triplet is also visible, similar to the Ca\,{\sc ii} absorption, at orbital phase 0.01 in the +73 days spectrum and possibly in the +104 days spectrum taken at phase 0.17. In the May spectrum, however, the (putative) Mg\,{\sc i} absorption is weaker and red-shifted by about $\sim$150\,km/s with respect to the binary rest frame, in rough  agreement with the radial velocity of the secondary star as determined by Thoroughgood et al. (2001). The eclipse spectrum suggests a systemic velocity of $\sim$50$\pm$9\,km/s which is also in agreement with the Thoroughgood et al. (2001) estimate.

The presence of the G-band suggests that the secondary star spectral type is not earlier than F3; while on the basis of the Mg\,{\sc i} morphology it is not later than G (Gray and Corbally 2009). 
We do not detect other convincing absorption lines or molecular bands at any wavelength. This implies that the U\,Sco secondary is not a giant star, and it points to a relatively hot donor, most likely hotter than K. However, it must be said that two facts limits our ability to detect other spectral features: 1) the relatively low signal-to-noise ratio of the NIR bands (SNR$\sim$3-12 depending on the wavelength and the portion of spectrum considered), and 2) the presence of emission components 
which may fill secondary star absorption lines. 
A secondary star of spectral type G or F should show a relatively strong Ca\,{\sc ii}(2) triplet at $\sim$8500-8600\AA, and at least the H Brackett\,$\gamma$ at 2.165\,$\mu$m. None of these are detected. The H-Paschen emission lines fill or partially fill the nearby Ca\,{\sc ii} absorption; 
 while the H-Brackett lines in emission at 
$\sim$1.736, 1.640 and 1.611\,$\mu$m might imply that the Brackett\,$\gamma$ emission is filling the secondary star Brackett\,$\gamma$ absorption. Dwarfs or sub-giants of late F or G spectral type have fairly shallow/weak Brackett absorption lines which can easily be ``filled'' by emission components at the same wavelength. 
At the same time we cannot confidently exclude an early K spectral type for U\,Sco secondary because we do not reach  (with reasonable signal to noise ratio) the wavelengths beyond 2.3\,$\mu$m, where the CO bands become visible. 

We note that the prime signature of an active stellar chromosphere in a G to M stars is the detection of emission cores or emission lines at the Ca\,{\sc ii}\,(1) and (2), H$\alpha$, Mg\,{\sc i}\,(2) and the Na\,{\sc i}\,D wavelengths (Gray and Corbally 2009). As we observe the Ca\,K\&H doublet and the Mg\,{\sc i}\,b triplet in absorption, and  because very active stars show the H-Balmer series in emission when Ca\,K\&H also appear in emission (see Gray et al. 2003) we believe the case of chromosphere activity is not applicable to U\,Sco. The non-detection of the Na\,{\sc i}\,D doublet of stellar origin can be explained if emission fills the absorption, either in the ejecta (e.g. the broad [N\,{\sc ii}]$\lambda$5755) or in the accretion disk (Na\,{\sc i}). Schaefer et al. (2011) report that the accretion disk size is of the order of 2.2\,R$_\odot\geq$R$_2$\footnote{Following Thoroughgood et al. (2001) system parameters determination.} from day +56 on. Therefore, it might well be that the cooler edges of the accretion disk remain visible during the secondary inferior conjunction, thus explaining the weakness of several H lines and the Na\,{\sc i}\,D lines.  This is even more likely if the U\,Sco accretion disk has raised edges downstream from the hot spot which remain uneclipsed at the inclination of $\sim$83$^o$ of the binary (see Section\,4, as well as Hachisu et al. 2000).

   \begin{figure*}[h!]
   \centering
   \includegraphics[width=6cm]{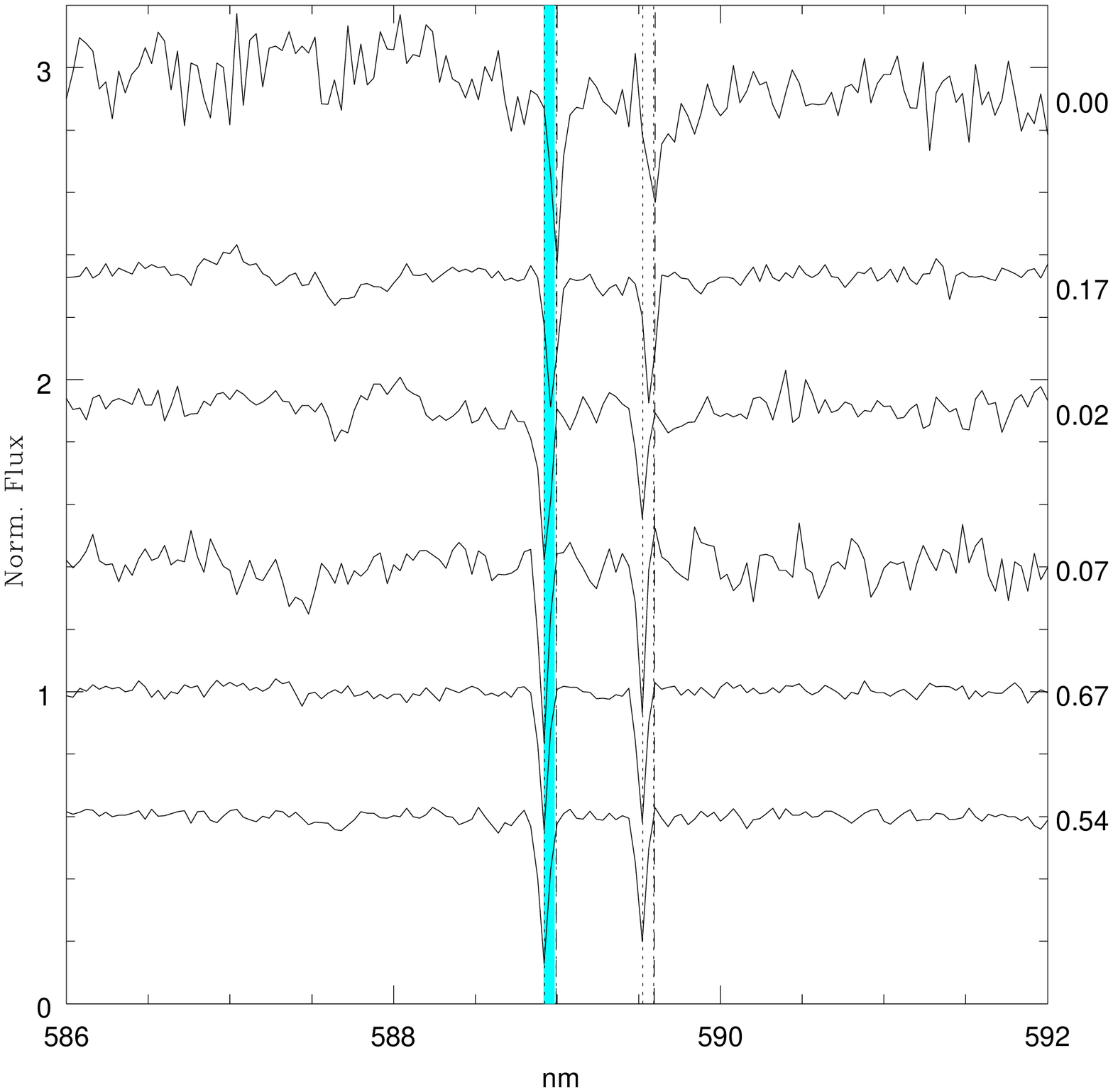}
   \includegraphics[width=6cm]{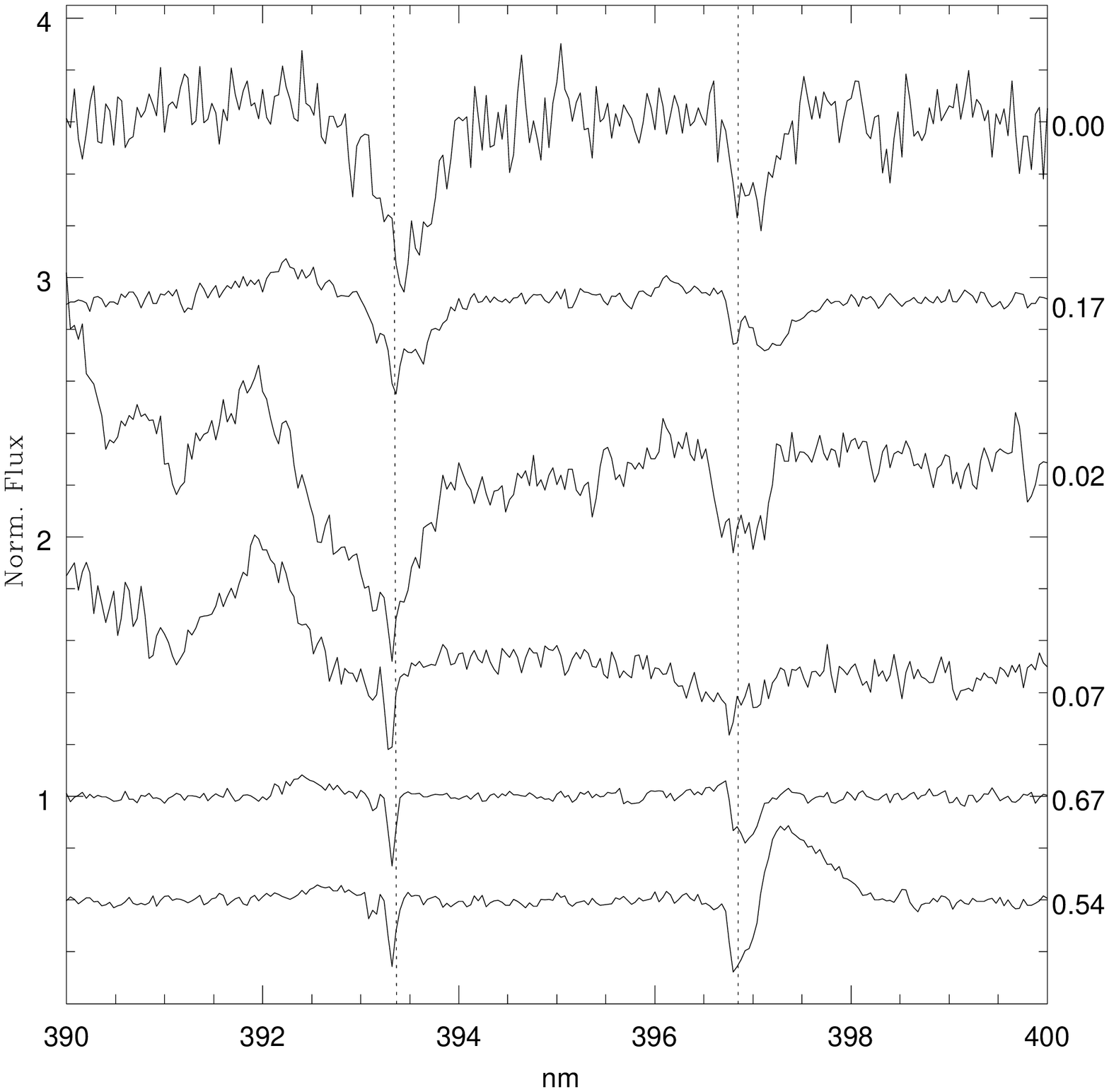}
   \includegraphics[width=6cm]{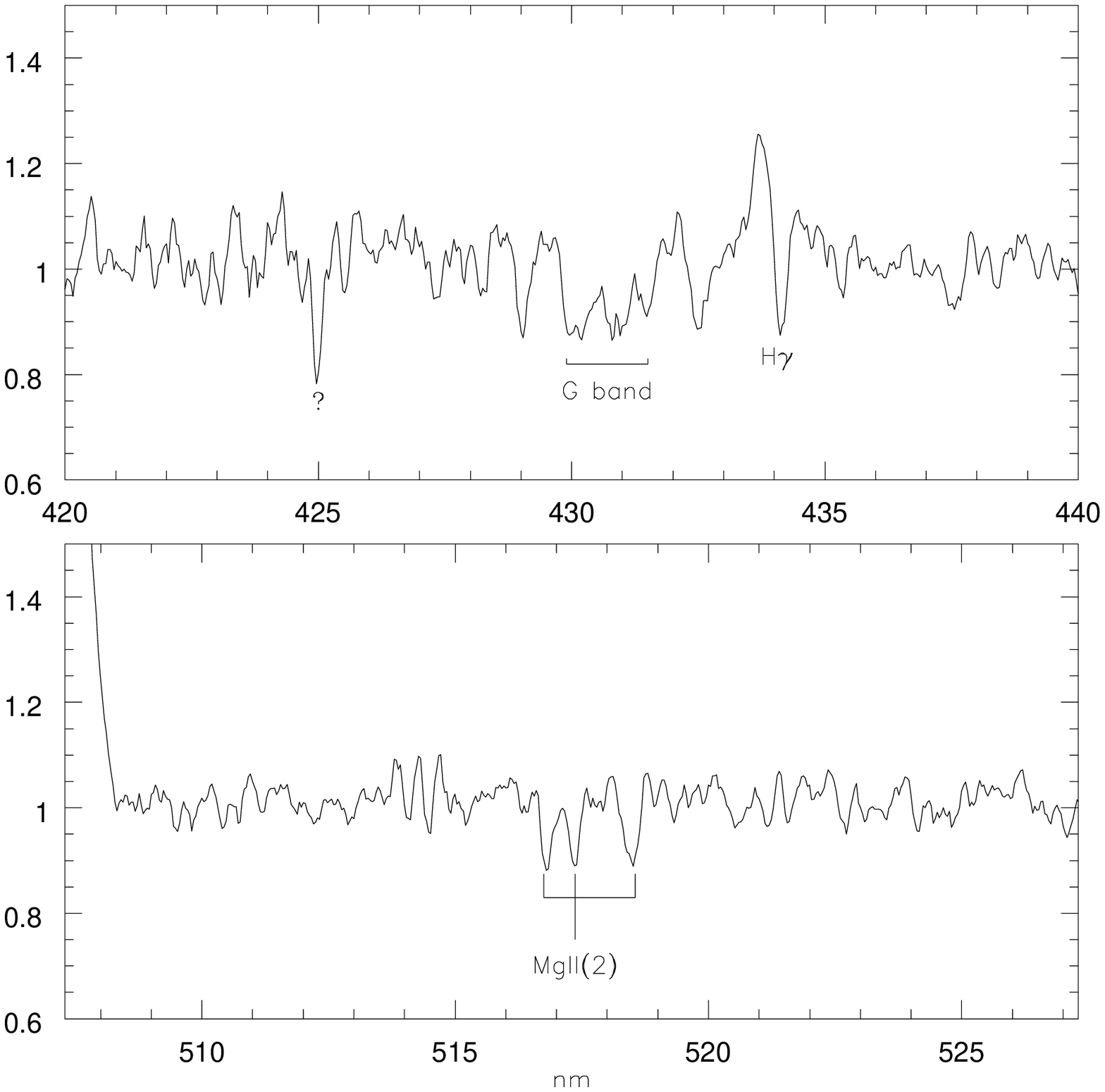}
   \caption{Left Panel: montage of X-Shooter spectra at the Na{\sc i} wavelength; the orbital phase corresponding to each spectrum is marked on the right side of the figure. The vertical dashed line marks the rest wavelength of the doublet, while the cyan-shaded area marks the drift observed in U\,Sco and the telluric spectra. Central Panel: montage of X-Shooter spectra at the Ca{\sc ii} wavelength. The vertical dashed lines marks the doublet rest wavelengths. Right Panel: selected portions of the July 2011 U\,Sco spectrum  showing the secondary star absorption features: the G band at $\sim$4300\AA\, (top), and the Mg\,{\sc i} triplet (bottom). 
}
              \label{}%
    \end{figure*}

\section{Discussion}

As has been observed in the past, U\,Sco outbursts develop very broad emission lines from H, He\,{\sc i} and N\,{\sc ii} (FWHM$\sim$8000\,km/s). These ejecta emission lines fade faster than the continuum and are superposed by a narrow emission component. The narrow component emits in H, He\,{\sc i} and {\sc ii}, and in the case of the He\,{\sc ii} lines do not have a corresponding broad component.  We have shown in this paper that this narrow component arises from the binary system (most likely the restored accretion flow) and is not part of the ejecta. In addition, the He\,{\sc ii} decrement is not optically thin but thick. Hence, a first strong implication is that the narrow emission lines and the He\,{\sc ii} should not be used to compute the abundance of the companion star, assuming recombination. Collisions appear to dominate the line formation mechanism in the newly formed accretion disk.

U\,Sco during the SSS phase and in quiescence is more similar to SSS sources and low-mass X-ray binaries (LMXB) than to nova-like and old-nova systems.  
For example, U\,Sco shows 1) strong H-Balmer and He-Pickering emission lines; 2) the He\,{\sc ii}$\lambda$4686 emission is stronger than the Balmer lines; 3) He\,{\sc ii} makes a substantial contribution to H emission; 4) H and He\,{\sc i}  absorption components distort the emission line profiles making it difficult to resolve the stars orbital motion. These absorption features are strong and well defined in the case of the H lines, but weak and often appearing as a steep edge  of the emission component in the case of the He emissions (e.g. RX\,J0019.8+2156, CAL\,87 and SMC\,13 in Cowley et al. 1998). 
In SSS the emission lines are associated to the accretion disk and their strength correlates with the absolute magnitude of the system (hence the accretion rate), more than L$_X$ (Cowley et al. 1998). The absorptions originate on the primary side, though they cannot be identified with the white dwarf. It is expected (Cowley et al. 1998 and reference therein) that SSS have a raised accretion disk edge on the side of the hot-spot, which extends half-way around the accretion disk, its actual extension being proportional to the accretion rate. Unlike SSS sources U\,Sco does not show O\,{\sc vi}\,$\lambda$3811 and C\,{\sc iv}$\lambda\lambda$5801,5811 emission lines. 

To continue the analogy between U\,Sco and the SSS, we note that a few of the SSS show variable and narrow satellite emission lines which have been explained by collimated jets (Cowley et al. 1998, Seward and Charles 2010). These have also been found to correlate with the accretion (high or low) state of the system and to possibly vary with the precession of a warped accretion disk (Cowley et al. 1998 and reference therein).  U\,Sco, in our picture of low mass transfer rate SSS is not expected to show jet-like emissions. However, during the SSS phase of the 1999 outburst, U\,Sco has displayed the tripled peaked profiles (Bonifacio et al. 1999, Munari et al. 1999, and Iijima  2002), which Kato and Hachisu (2003) have explained with a $\sim$5$^o$ collimated jet (but see also below).  
Several SSS sources, likewise U\,Sco, seem to have a binary ratio $q=$M$_2$/M$_1<$1 contrary to theoretical predictions (van\,den\,Heuvel et al. 1992) and the accretion disk outshining the secondary star.  

Most of the above characteristics are common to LMXBs, e.g Sco\,X-1  (Lewing et al. 1995; Steeghs \& Casares 2002) and MM\,Ser (Hynes et al. 2004). 
However, LMXB emission lines  arise primarily from the irradiated face of the secondary star (see, for example the Doppler maps produced for Sco\,X-1 by Steeghs and Casares 2002) or are produced by fluorescence. Conversely, no fluorescence emission nor significant emission lines from the irradiated secondary  are observed in U\,Sco  (see e.g. Fig.6 and the Doppler tomographs produced by Thoroughgood et al. 2001). 

Given all of the above, it would be interesting to monitor  U\,Sco well into quiescence in order to ascertain whether those similarities persist at that phase too, or are instead limited to the post-outburst epochs. 

It is important to note that U\,Sco accretion resumes as early as $\sim$8-10 days after maximum and this is not peculiar to the 2010 outburst. 
Our sequence of spectra has shown that the accretion re-starts on day +8 with the appearance of the narrow emission components.  Despite their lower resolution, the He\,{\sc ii} narrow component at 4686\AA\, is visible in the +7.2 day NIR spectrum of Evans et al. (2001), and in the +11 days spectrum of Iijima (2002). The spectrum of Iijima (2002), shows, in addition, that on day +11 Bowen fluorescence is still active. Also, the spectrum of Sekiguchi et al. (1988) on day +9 and the +8 day spectrum of Barlow et al. (1981) show narrow emission components and the first appearance of He\,{\sc ii}$\lambda$4686. 
It is therefore tempting to conclude that this is evidence for the resumption of accretion in the early decline spectra of the 1999, 1987, and 1979 outbursts. 
Unfortunately, most of the past follow-ups did not monitor the modulation and changes of the narrow emission component with the binary orbital phase. However, Barlow et al. (1981) observe that both the He\,{\sc ii}$\lambda$4686 and the Balmer emission lines profiles change significantly from day +8 to +12; while their Figure\,3 shows that the peak position of the H$\gamma$ emission line changes with time and is split by a relatively strong absorption component at orbital phase 0.72, in agreement with our observations (Fig.\,5). Munari et al. (1999) present a sequence of four spectra centered on H$\alpha$ and taken between days $\sim$+19 and +22, while U\,Sco was displaying a triple peaked profile. This sequence includes two low resolution spectra and two spectra taken at the same orbital phase. Therefore, it difficult to disentangle differences from instrumental effects that are possibly intrinsic to the system. Inspection of Figure\,3 and 4 of Munari et al. (1999) shows a change in the relative intensity and profile of the three emission components, as well as a decrease in the separation of the peaks with time.  
Combining our results from the U\,Sco 2010 outburst with the observations of Barlow et al. (1981) and Munari et al. (1999) we conclude that the U\,Sco secondary star always resumes mass transfer $\sim$8-10 days after the outburst. 

An alternative explanation for the partially collimated jets proposed by Kato and Hachisu (2003) would be to interpret the satellite emission at $\pm$1500\,km/s as due to the restored accretion. The intensity of the satellite emission in the 1999 outburst\footnote{Note, however, that Barlow et al. (1981), who also reported the development of the triple peaked emission with similar velocities, observe much smaller intensities in their spectra taken at day +12 and +18.} and their rapidly decreasing separation are hardly compatible with jets, which typically change in profile and positions (possibly because of a precessing disk) on time scales of months to years (see Crampton et al. 1996; Cowley et al. 1998).  Conversely, velocities of $\pm$1500\,km/s are possibly consistent with accretion columns in the presence of a magnetic field, or Keplerian velocities at about 10 white dwarf radii. Independent of their exact origin, the U\,Sco satellite emission forms only after accretion has resumed, implying that its formation depends on the mass transfer rate at the onset of the accretion and possibly on and the geometrical configuration of the system (/of the magnetic poles), should U\,Sco host a magnetic white dwarf. 

We note that U\,Sco accretion and soft X-ray emission are coeval, implying that the  SSS phase is fueled at least in part by the recovered accretion. Therefore, in the case of U\,Sco it is not appropriate to conclude that the white dwarf ejects less mass than it accretes over an outburst cycle on the basis of the observed SSS phase. 
It will be important to verify how many of these characteristics are common to other recurrent novae of the same type, as well as monitor and characterize the resumption of the accretion process in other classical and recurrent novae.

\section{Summary and conclusions}

The analysis of the data presented here shows the following results
\begin{itemize}
\item In U\,Sco, the mass transfer from the secondary star recovers $\sim$8-10 days after maximum at every outburst. The narrow emission lines from H, He\,{\sc i} and He\,{\sc ii} lines arise from the optically thick gas of the reforming accretion disk  
\item The SSS phase (plateau phase in the light curve) is at least in part fueled by accretion and not just by residual activity on the surface of the white dwarf. 
\item The ejecta do not produce He\,{\sc ii} emission lines. Only weak He\,{\sc i} emission lines form in the ejecta and they are primarily blended with transitions from low ionization elements, thus preventing a reliable abundance calculation.   
The narrow He\,{\sc i} and He\,{\sc ii} emission lines from the disk, being optically thick, do not allow abundance calculation either.  
\item The secondary star spectral type remains ambiguous though constrained between a F3 and a G sub-giant. Higher signal to noise data taken during the eclipses are needed in order to better identify the donor spectral type and ascertain which absorption features show emission components due to a Wilson-Bappu effect.
\item  U\,Sco spectral characteristics during the post outburst phases are remarkably similar to those of SSS object and LMXBs and might deserve further attention. 
\end{itemize}

\begin{acknowledgements}
E.M. is thankful to Prof B.E. Schaefer for having shared his photometric data, and to Prof. A. Bianchini for the stimulating discussions on some of the idea presented in this paper. The authors are thankful to the ESO Director General for having approved the DDT proposal for which the X-Shooter and part of the FEROS spectra were obtained. 
\end{acknowledgements}

\begin{appendix} 
\section{}

\begin{figure*}
\centering
\includegraphics[width=13cm, angle=270]{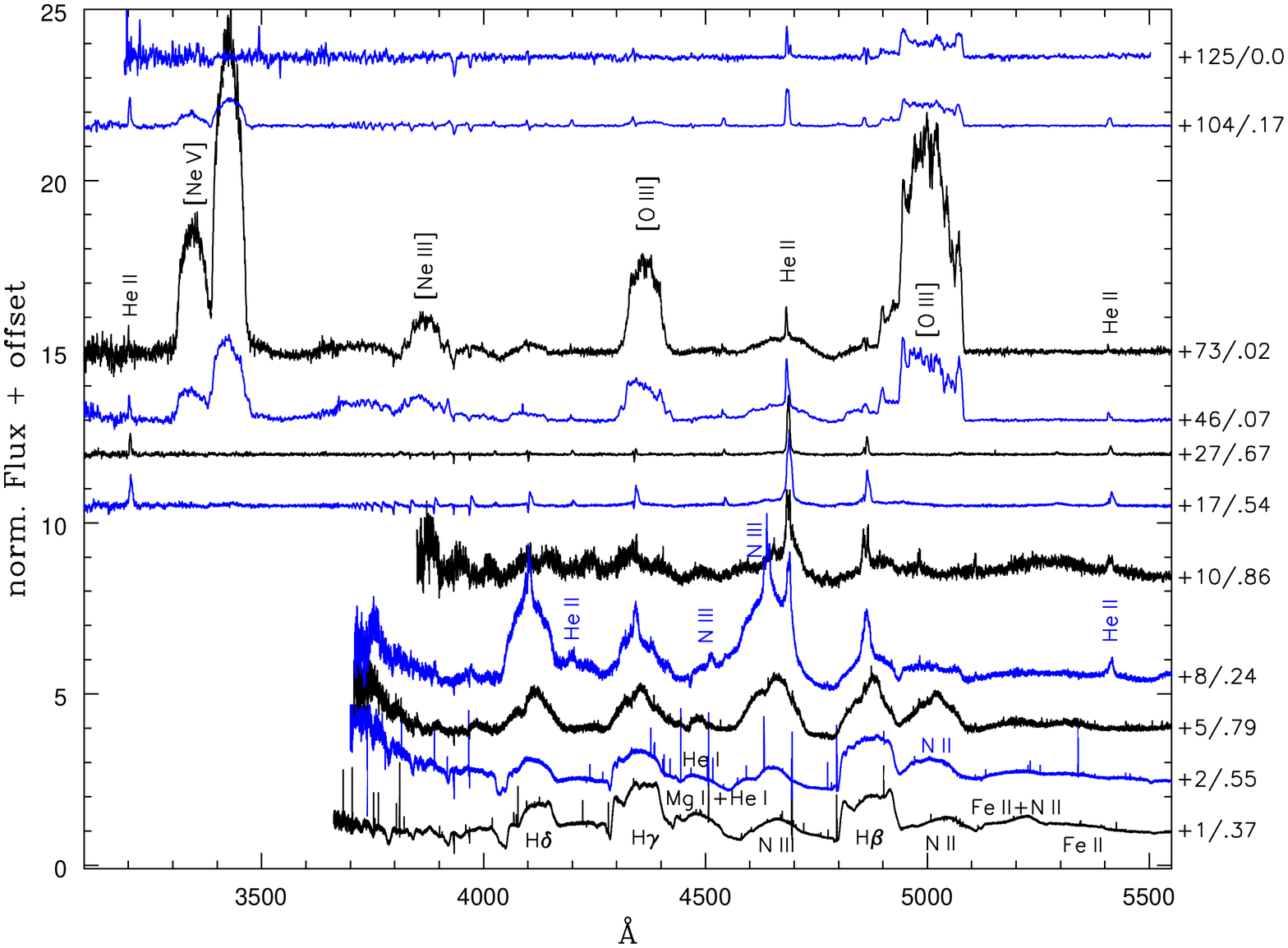}
\caption{The spectral evolution (FEROS and X-Shooter spectra) of U\,Sco from day +1 to +125 after maximum  in the wavelength range matching X-Shooter UVB arm. The notations on the right side of the panel mark the ``age'' of the spectrum (day after maximum) and the binary orbital phase. The alternating black and blue color are just for clarity. }
\label{VIS}%
\end{figure*}

\begin{figure*}
\centering
\includegraphics[width=13cm, angle=270]{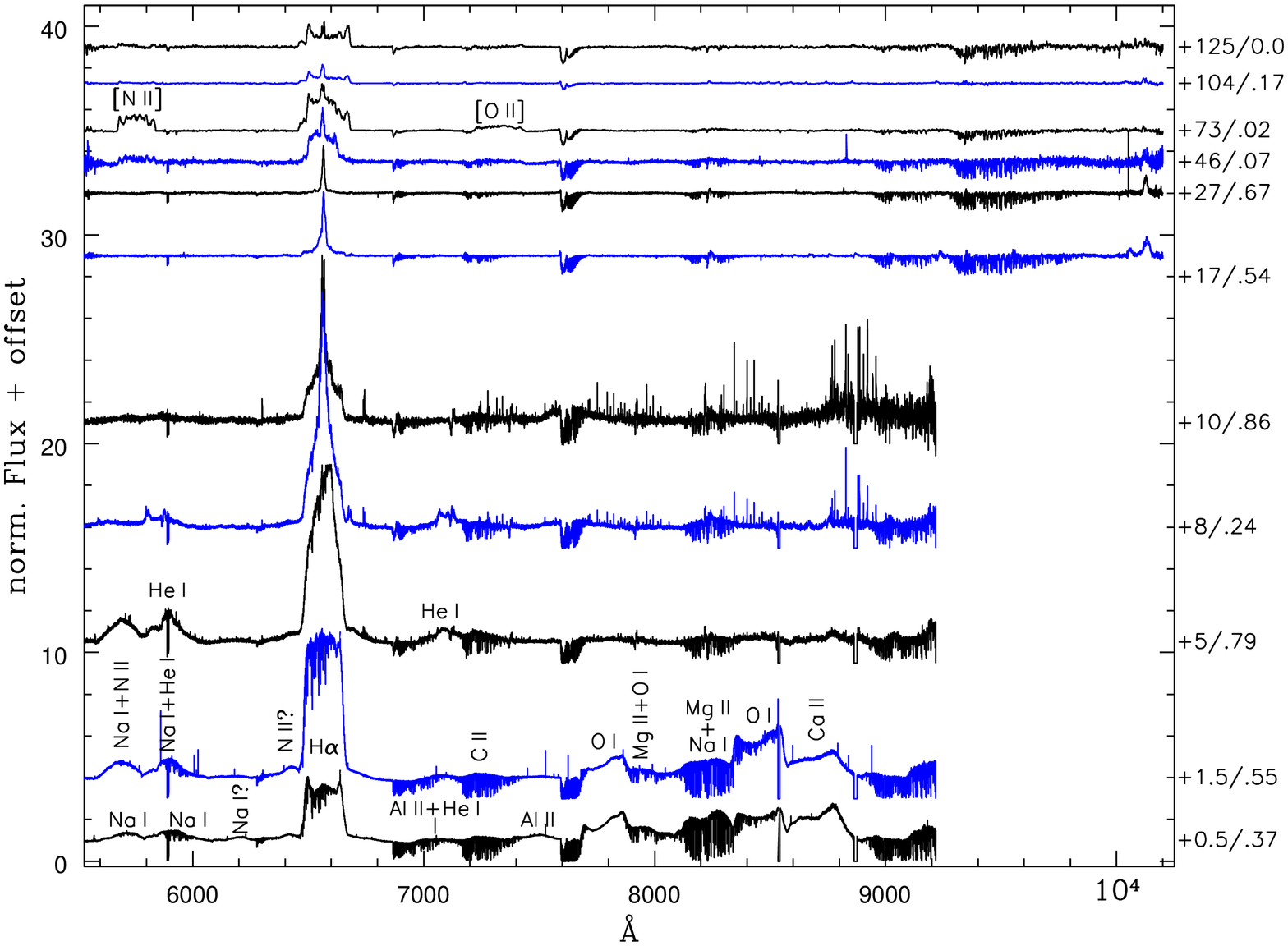}
\caption{The spectral evolution (FEROS and X-Shooter spectra) of U\,Sco from day +1 to +125 after maximum  in the wavelength range matching X-Shooter VIS arm. The notations on the right side of the panel mark the ``age'' of the spectrum (day after maximum) and the binary orbital phase. The alternating black and blue color are just for clarity.}
\label{NIR}%
\end{figure*}

\begin{figure*}
\centering
\includegraphics[width=13cm, angle=270]{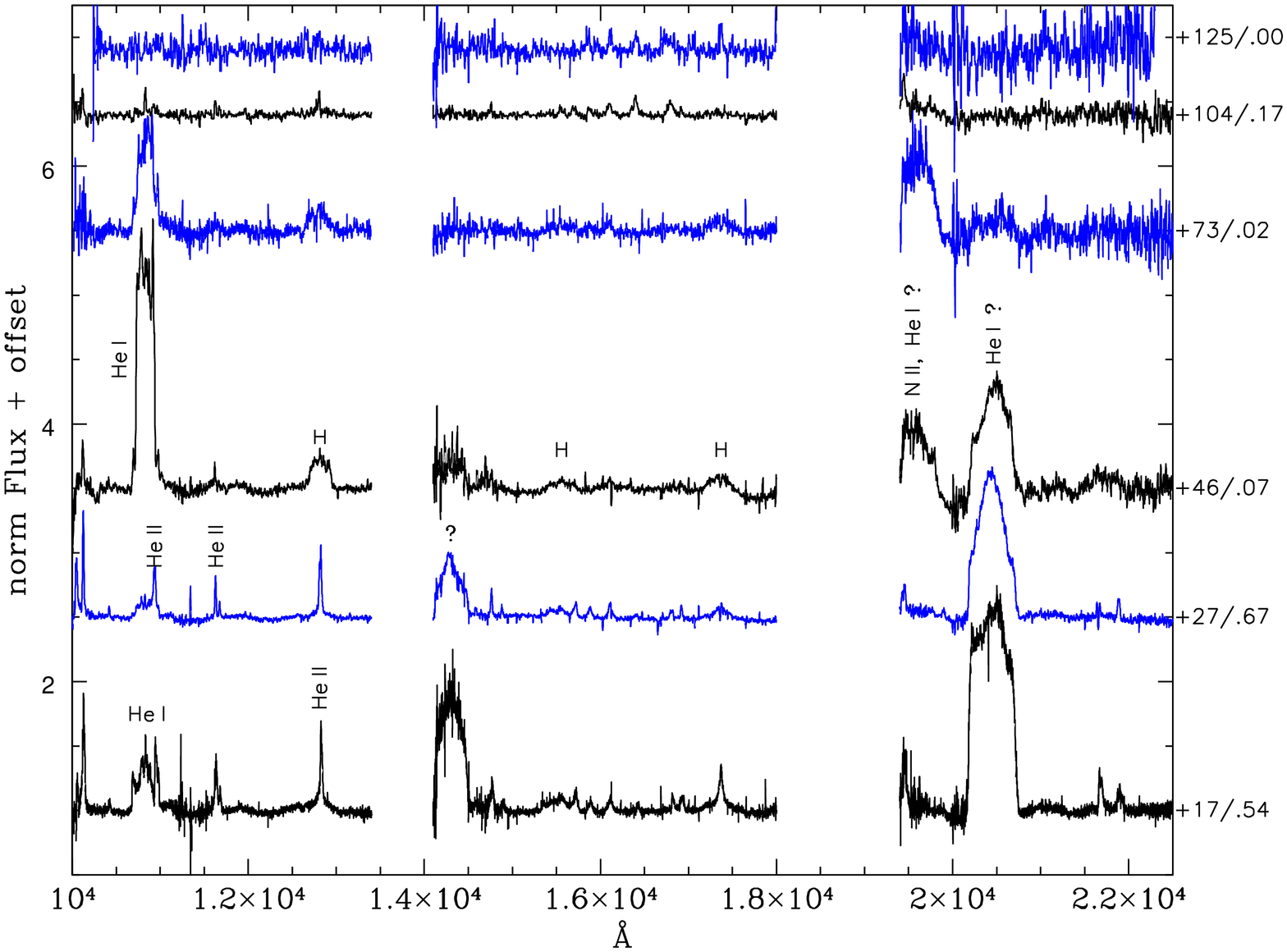}
\caption{The spectral evolution (X-Shooter spectra) of U\,Sco from day +17 to +125 after maximum  in the wavelength range matching X-Shooter NIR arm. The notations on the right side of the panel mark the ``age'' of the spectrum (day after maximum) and the binary orbital phase. The alternating black and blue color are just for clarity.}
\label{UVB}%
\end{figure*}

\begin{table*}[t]
 \centering
\scriptsize
\flushleft
 \begin{minipage}{140mm}
  \caption{Integrated flux of the ejecta (broad components) during the early decline (FEROS spectra). The line fluxes are in units of H$\beta$ (which flux, in erg/sec/cm$^2$/A, is given between brackets) and have been measured on the dereddened spectra, assuming E(B-V)=0.15 and R=3.1, scaled to the nova V magnitude. Note that the Feb\,7 has not been included as the SNR is too low for all the lines but H$\alpha$. }
  \begin{tabular}{@{}lcccc@{}}
  \hline
ID & Jan\,29 & Jan\,30 & Feb\,2 & Feb\,5  \\
\hline
H$\varepsilon$+Ca{\sc ii} & 0.57 & 0.51 & - & - \\
H$\delta$ & 0.65 & 0.54 & 1.08(+N{\sc iii(1)}?) & 3.45 \\
H$\gamma$ & 0.74 & 0.54(+He{\sc i}?) & 0.78 & 0.98 \\
He{\sc i}$\lambda$4471+Mg{\sc ii} & 0.39 & 0.31 & 0.28 & - \\
BowenBlend & 0.51 & 0.59 & 1.79 & 4.50 \\
H$\beta$ & 1/(6.77e-11) & 1/(4.85e-11) & 1/(1.92e-11) & 1/(5.22e-12) \\
N{\sc ii(64)} & 0.30 & 0.40 & 0.61 & 0.59 \\
Fe{\sc ii(42)}+N{\sc ii(66)} & 0.84 & - & - & - \\
Na{\sc i (6)}+N{\sc ii(3)} & 0.22 & 0.29 & 0.44 (N{\sc ii}) & 0.27 \\
Na{\sc i}D+He{\sc i} & 0.22 & 0.23 & 0.47 (He{\sc i}) & 0.23 \\
Na{\sc i}? & 0.09 & 0.036 & 0.06  & -\\
N{\sc ii(46)}? & 0.16 & 0.11 & - & - \\
H$\alpha$ & 2.33 & 3.36 & 3 & 2.53 \\
Al{\sc ii(3)}+He{\sc i} & 0.10 & 0.10 (He{\sc i}) & 0.21 & 0.21 \\  
C{\sc ii(3)} & 0.09: & $>$0.007 & $>$0.08 & - \\
Al{\sc ii}(21) & 0.09 & 0.017 & - & - \\
O{\sc i(1)} & 0.53 & 0.24 & - & - \\
O{\sc i(64)}+Mg{\sc ii}(8) & 0.14 & 0.05 & - & -\\
Mg{\sc ii(7)}+Na{\sc i} & 0.40  & 0.12 & 0.04 & -\\
O{\sc i(4)} & 0.74 & 0.66 & 0.12 & -\\
Ca{\sc ii}(2) & 0.81 & 0.29 & - & -\\
\hline
\end{tabular}
\end{minipage}
\end{table*}

\begin{table*}[t]
 \centering
\scriptsize
\flushleft
 \begin{minipage}{140mm}
  \caption{Integrated flux of the ejecta (broad components) during the late decline (X-Shooter spectra). The line fluxes are in units of H$\beta$ (which flux, in erg/sec/cm$^2$/A, is given between brackets) for all epochs but Feb\,24, and have been measured on the dereddened spectra, assuming E(B-V)=0.15 and R=3.1, scaled to the nova V magnitude.  }
  \begin{tabular}{@{}lcccccc@{}}
  \hline
ID & Feb\,14 & Feb\,24 & Mar\,15 & Apr\,11 & May\,12 & Jul\,5  \\
\hline
$[$Ne{\sc v}$]$(1) & - & - & 2.44 & 6.16 & 14.5 & - \\
$[$Ne{\sc v}$]$(1) & - & - & 6.91 & 16.7 & 42.2 & - \\
$[$Ne{\sc iii}$]$(1) & - & - & 1.79 & 3.59 & - & - \\
$[$Ne{\sc iii}$]$(1) & - & - & 0.42 & 0.92 & - & - \\
H$\delta$ & - & - & 0.65 & 0.77 & - & - \\
$[$O{\sc iii}$]$(2) & - & - & 3.41 & 6.05 & 6.58 & - \\
BowenBlend & 2.09 & - & 2.11 & 3.06 & - & - \\
H$\beta$ & 1/(3.83e-14) & - & 1/(4.92e-15) & 1/(1.77e-15) & 1/(1.16e-16) & 1/(7.1e-17) \\
$[$O{\sc iii}$]$(1) & - & - & 8.26 & 24.6 & 64.83 & 31.8 \\
$[$N{\sc ii}$]$(3) & - & - & 1.01 & 2.86 & 4.41 & - \\
H$\alpha$ & 1.62 & 1/(2.20e-14) & 3.35 & 6.61(+[N{\sc ii}](1)) & 34.22(+[N{\sc ii}](1))  & 47.6(+[N{\sc ii}](1)) \\
7316 $[$O{\sc ii}$]$(2) & - & - & 0.36 & 1.18 & - & - \\
He{\sc i} & 1.37(+H) & 0.34 & 2.20 & 1.79 & - & - \\
12810 H$\lambda$12818 & - & - & 0.18 & - & - & - \\
$\sim$14297 -- N{\sc ii}$\lambda$14287? or O{\sc iii}? & 2.07 & 0.76 & - & - & - & - \\
19571 -- He{\sc i}? & - & - & 0.23 & - & - & - \\
20440 -- He{\sc i}$\lambda$20424? & 2.77 & 1.01 & 0.44 & - & - & - \\
\hline
\end{tabular}
\end{minipage}
\end{table*}

\begin{table*}[t]
\centering
\scriptsize
\flushleft
 \begin{minipage}{140mm}
  \caption{Integrated flux of the He\,{\sc ii} narrow emissions and flux ratio with respect to the He\,{\sc ii}\,$\lambda$4686. Only those lines which are not blending with H have been listed in the table.  Flux (in erg/sec/cm$^2$/A) have been measured on the dereddened spectra, assuming E(B-V)=0.15 and R=3.1.  }
  \begin{tabular}{@{}lcccccc@{}}
  \hline
ID & Feb\,14 & Feb\,24 & Mar\,15 & Apr\,11 & May\,12 & Jul\,5 \\
\hline
3203  & 5.008E-14/0.622 & 1.468E-14/0.481 & 6.341E-16/0.413 & 9.595E-17/0.347 & 4.021E-16/0.423 & \\
4200  & 4.191E-15/0.052 & 1.774E-15/0.058 & 1.437E-16/0.093 & 3.330E-17/0.121 & 1.074E-16/0.113 & \\
4542  & 5.454E-15/0.068 & 2.006E-15/0.066 & 1.912E-16/0.124 & 3.702E-17/0.134 & 1.508E-16/0.159 & \\
4686  & 8.049E-14/1.000 & 3.055E-14/1.000 & 1.537E-15/1.000 & 2.761E-16/1.000 & 9.510E-16/1.000 & \\
5411  & 1.084E-14/0.135 & 2.927E-15/0.096 & 2.032E-16/0.132 & 3.656E-17/0.132 & 1.796E-16/0.189 & \\
10123 & 1.903E-14/0.236 & 4.634E-15/0.151 & 6.507E-16:&           &  & \\
\hline
\end{tabular}
\end{minipage}
\end{table*}

\end{appendix}

\end{document}